\documentclass[aps,twocolumn,
showpacs,pra,amssymb, amsmath]{revtex4}
\usepackage{amsmath,latexsym,amssymb}
\usepackage[dvips]{graphicx}
\usepackage{amsmath}
\usepackage{latexsym}
\usepackage{amssymb}
\usepackage{bm}

\begin{document}

\title{Kochen-Specker 
theorem as a precondition for secure quantum key distribution}
\author{Koji Nagata}
\affiliation{National Institute of Information and Communications 
Technology, 4-2-1 Nukuikita, Koganei, Tokyo 184-8795, Japan}
\date{\today}

\begin{abstract}

We show that (1) the violation of the Ekert 91 inequality is a 
sufficient condition for certification of the Kochen-Specker (KS) theorem, 
and (2) the violation 
of the Bennett-Brassard-Mermin 92 (BBM) inequality is, also, 
a sufficient condition for certification of
the KS theorem.
Therefore the success in each QKD protocol 
reveals the nonclassical feature of quantum theory,
in the sense that the KS realism is violated.
Further, it turned out that the Ekert inequality and the BBM inequality are 
depictured by distillable entanglement 
witness inequalities. Here, we connect the 
success in these
two key distribution processes into the no-hidden-variables theorem
 and into witness on distillable entanglement.
We also discuss the explicit difference between 
the KS realism and Bell's local realism
in the Hilbert space formalism of quantum theory.
\end{abstract}

\maketitle

Quantum key distribution (QKD) allows the participants (Alice and Bob) 
to wish to agree on 
a secret key about which no eavesdropper 
can obtain significant information, by the properties of quantum 
theory \cite{bib:Shor}.
There are many researches related to the security of 
QKD schemes \cite{bib:Shor,bib:Koashi,bib:Acin1,bib:Curty,bib:Acin2}.
Especially, the connection between entanglement, 
which is a resource of various quantum information 
processes \cite{bib:Nielsen,bib:Galindo}, 
and the success in QKD 
has been discussed flourishingly \cite{bib:Acin1,bib:Curty,bib:Acin2}.

As for the connection between no-hidden-variables theorem
and the success in QKD, 
in 1991, Ekert proposed \cite{bib:Ekert} a 
QKD protocol based on Bell's theorem \cite{bib:Bell}
using Einstein-Podolsky-Rosen (EPR) correlation.
Ekert discussed the success in his QKD process 
in cryptography depends on the completeness of quantum theory by
invoking Bell's theorem.
However, in 1992, Bennett, Brassard and Mermin 
showed \cite{bib:Bennett} that 
Bell's theorem is not an essential part of successful QKD protocols.
They proposed a simple QKD protocol (BBM protocol) 
using the EPR correlation, 
without invoking Bell's theorem to detect an eavesdropper's listening.
And they discussed that the 
BBM protocol is, essentially, equivalent to 
the well known BB84 scheme.
They concluded that the EPR effect is superficial for 
successful QKD protocols by
invoking the BB84 protocol, which uses single particles 
instead of the EPR correlation.

One can see the success in each protocol is certified by 
violations of their (Ekert and BBM) inequalities, respectively, 
because of coexistence with the recent 
researches \cite{bib:Acin1,bib:Curty,bib:Acin2}.
To explain detail, we shall introduce a quantum state
which has convenient property.
Please note that there is a quantum state, 
on which simulates the actual measured random bits by single-copy 
measurements \cite{bib:Acin1,bib:Acin2}.
We shall call such a state the {\it candidate} state for short.
One will then see the success in each of EPR-type-QKD protocols 
(Ekert 91 and BBM 92), 
in terms of violations of their inequalities,
implies that the candidate state can not 
be a separable state.
The candidate state
must have the fidelity (more than half) to the EPR state.
Thereby, the candidate state is necessarily 
a distillable entangled state in two qubits systems \cite{bib:Nagata3}.
This point coexists with recent arguments
in \cite{bib:Acin1,bib:Curty,bib:Acin2}.

In this paper, we connect the success in each 
protocol (Ekert 91 and BBM 92)
into the no-hidden-variables theorem, using the EPR effect.
We show that (1) the success in the Ekert protocol reveals 
the nonclassical property of quantum theory, 
and (2) the success in the BBM protocol, also, 
reveals the nonclassical property of quantum theory.
One can see the nonclassical property of 
quantum theory is certified by 
the Kochen-Specker (KS) theorem \cite{bib:KS} 
as well as Bell's theorem \cite{bib:com}. 

We will see 
the simple BBM protocol can be seen 
to be equivalent to 
the Ekert protocol in the sense that the criterion of the success 
depends on the fidelity to the EPR correlation.
That is, the fidelity to the EPR correlation, which is more than half, 
gives secret-key distillability.
This allows explicit construction of inequalities valid for 
the KS realism.
The violation of the KS type of realism is a necessary 
condition for certification of the success in each of QKD schemes
(Ekert 91 and BBM 92), introducing the candidate state.
Thereby, one connects the success in each of
 QKD protocols 
into the no-hidden-variables theorem.
Finally, we shall discuss the situations in which 
the violation of their (Ekert and BBM) inequalities 
in each of QKD protocols does not invalidate 
Bell's local realism.

Let us start analyzing the Ekert inequality and the BBM inequality.
Let $\rho$ be 
the candidate state which simulates 
the actual measured correlation by single-copy of itself.
The Ekert inequality is as follows:
\begin{eqnarray}
|E_{\rho}(a_1 b_1)-E_{\rho}(a_1 b_3)
+E_{\rho}(a_3 b_1)+E_{\rho}(a_3 b_3)|\leq \sqrt{2}.\label{Ekert}
\end{eqnarray}
Here, $E_{\rho}(o)$ represents an expectation value 
($E_{\rho}(o)=tr[\rho o]$, $o$ is a quantum observable).
Ekert chose the measurement observables as
$a_1=\sigma^a_x$, 
$a_3=\sigma^a_y$,
$b_1=\frac{\sigma^b_y+\sigma^b_x}{\sqrt{2}}$, and
$b_3=\frac{\sigma^b_y-\sigma^b_x}{\sqrt{2}}$,
where $\sigma^a_l$ and $\sigma^b_l\   (l=x, y, z)$ 
are the Pauli observables with respect to Alice and 
Bob, respectively.
Hence, (\ref{Ekert}) is rewritten by
\begin{eqnarray}
|E_{\rho}(\sigma_x^a \sigma^b_x)+E_{\rho}(\sigma^a_y \sigma^b_y)|\leq 1.
\end{eqnarray}
We shall show that the violation of this inequality implies that 
the candidate state 
has the fidelity to the EPR state, which is more than half.
We notice that 
$
tr[\rho \sigma_x^a\sigma^b_x]+tr[\rho\sigma^a_y\sigma^b_y]
=2tr[\rho 
(|\Psi^+\rangle\langle\Psi^+|-|\Psi^-\rangle\langle\Psi^-|)]
$
with
$|\Psi^+\rangle=\frac{|+^a-^b\rangle+|-^a+^b\rangle}{\sqrt{2}}$ and
$|\Psi^-\rangle=\frac{|+^a-^b\rangle-|-^a+^b\rangle}{\sqrt{2}}$.
Here, $\sigma^a_z|\pm^a\rangle=\pm|\pm^a \rangle$ and 
$\sigma^b_z|\pm^b\rangle=\pm|\pm^b \rangle$.
Therefore, if (\ref{Ekert}) is violated,
we have
$\langle\Psi^+|\rho|\Psi^+\rangle>\frac{1}{2}$
or
$\langle\Psi^-|\rho|\Psi^-\rangle>\frac{1}{2}$.
Hence, the fidelity ($>1/2$) to the EPR state 
($|\Psi^+\rangle$ or $|\Psi^-\rangle$) is established if (\ref{Ekert})
is violated.
It is well known that this condition is sufficient to $\rho$ be a 
distillable
entangled state since $\rho$ is a two qubits state.

Let us derive the original Ekert's form written in \cite{bib:Ekert}.
Assume that the correlation is disturbed 
(or replaced) by eavesdroppers.
Imagine the candidate state $\rho$ can be any separable state written by
\begin{eqnarray}
\rho=\int p(n^a,n^b)dn^a dn^b \rho^a\otimes\rho^b,\label{separable}
\end{eqnarray}
where
$\rho^a=\frac{1}{2}(I+n^a\cdot\sigma^a)$ and
$\rho^b=\frac{1}{2}(I+n^b\cdot\sigma^b)$.
Here, $\sigma^a=(\sigma^a_x, \sigma^a_y, \sigma^a_z), 
\sigma^b=(\sigma^b_x, \sigma^b_y, \sigma^b_z)$ and $I$ represents 
the identity operator. 
The normalized measure $p(n^a,n^b)$ and unit vectors $n^a, n^b$
virtually represent
strategy of the eavesdroppers.
Let us introduce $\vec{a}_k$ and $\vec{b}_k\  (k=1, 3)$ such that 
$\vec{a}_k\cdot\sigma^a=a_k$ and $\vec{b}_k\cdot\sigma^b=b_k$.
Let $S$ be 
\begin{eqnarray}
S=tr[a_1 b_1\rho]-
tr[a_1 b_3\rho]+
tr[a_3 b_1\rho]+
tr[a_3 b_3\rho].\label{Sin}
\end{eqnarray}
Since
$\{a_1, a_3\}={\bf 0}$ and  
$\{b_1, b_3\}={\bf 0}$ hold,
the upper bound of $|S|$ is $\sqrt{2}$ \cite{bib:Nagata4} 
when $\rho$ is any separable state.
On substituting (\ref{separable}) into 
(\ref{Sin}) and performing some algebra we find that 
\begin{eqnarray}
&&S=\int p(n^a,n^b)dn^a dn^b\nonumber\\
&&((\vec{a}_1\cdot n^a)(\vec{b}_1\cdot n^b)-
(\vec{a}_1\cdot n^a)(\vec{b}_3\cdot n^b)\nonumber\\
&&+
(\vec{a}_3\cdot n^a)(\vec{b}_1\cdot n^b)+
(\vec{a}_3\cdot n^a)(\vec{b}_3\cdot n^b)).
\end{eqnarray}
This leads us to the original form of the Ekert inequality.
Thereby, we see that the 
Ekert inequality is a distillable entanglement witness inequality.

Let us move to the BBM
inequality, which is as follows:
\begin{eqnarray}
|E_{\rho}(\sigma^a_x \sigma^b_x)
+E_{\rho}(\sigma^a_z \sigma^b_z)|\leq 1.\label{BBM}
\end{eqnarray}
Immediately, we have
$tr[\rho \sigma_x^a\sigma^b_x]+tr[\rho\sigma^a_z \sigma^b_z]=2tr[\rho 
(|\Phi^+\rangle\langle\Phi^+|-|\Psi^-\rangle\langle\Psi^-|)]$.
Here,
$
|\Phi^+\rangle=\frac{|+^a+^b\rangle+|-^a-^b\rangle}{\sqrt{2}}
$.
Therefore, if (\ref{BBM}) is violated,
we have
$\langle\Phi^+|\rho|\Phi^+\rangle>\frac{1}{2}$
or
$\langle\Psi^-|\rho|\Psi^-\rangle>\frac{1}{2}$.
Hence, the fidelity ($>1/2$) to the EPR state
($|\Phi^+\rangle$ or $|\Psi^-\rangle$) is established.
Let us introduce $\vec{x}$ and $\vec{z}$ such that 
$\vec{x} \cdot\sigma^a=\sigma^a_x$,
$\vec{z}\cdot\sigma^a=\sigma^a_z$,
$\vec{x}\cdot\sigma^b=\sigma^b_x$, and
$\vec{z}\cdot\sigma^b=\sigma^b_z$.
Let $T$ be
\begin{eqnarray}
T=tr[\rho \sigma_x^a\sigma^b_x]+tr[\rho\sigma^a_z \sigma^b_z].\label{Tin}
\end{eqnarray}
Please notice the upper bound of $|T|$ is $1$ 
when $\rho$ is any separable state \cite{bib:Nagata5}.
On substituting (\ref{separable}) into 
(\ref{Tin}) and performing some algebra we find that 
\begin{eqnarray}
&&T=\int p(n^a,n^b)dn^a dn^b\nonumber\\
&&((\vec{x}\cdot n^a)(\vec{x}\cdot n^b)+
(\vec{z}\cdot n^a)(\vec{z}\cdot n^b)).
\end{eqnarray}
This leads us to the original form of the BBM inequality.
Thereby, we see that the BBM
inequality is a distillable entanglement witness inequality.

In what follows, we shall show that 
violations of (\ref{Ekert}) and of (\ref{BBM})
give explicit sufficient conditions to refute the KS type of realism.
The KS realism is equivalent the following situation:
all quantum observables commute simultaneously \cite{bib:com}.

Let us 
consider a classical probability space $(\Omega,\Sigma,\mu_{\rho})$, where
$\Omega$ is a nonempty sample space, $\Sigma$ is an algebra of subsets of $\Omega$, and $\mu_{\rho}$ is a 
normalized measure on $\Sigma$ such that 
$\mu_{\rho}(\Omega)=1$.
The subscript $\rho$ expresses the following meaning:
the probability measure $\mu_{\rho}$ is determined uniquely
when a candidate state $\rho$ is specified.

Let us introduce realistic functions (classical random variables) 
$f(o,\omega)$
onto $\Omega$.
Here $o$ is an observable and $\omega\in\Omega$ is a hidden variable.
One may assume that the possible values of $f(o,\omega)$ 
are eigenvalues of corresponding observable $o$.
One may, also,  
assume the probability measure $\mu_{\rho}$ is chosen such that the following relation
is valid:
\begin{eqnarray}
tr[\rho o]=E_{\rho}(o)=\int_{\omega\in\Omega}\mu_{\rho}(d\omega)
f(o,\omega)
\end{eqnarray}
for every observable $o$.
Please notice these two assumptions 
for the realistic functions $f$ and for the probability measure $\mu_{\rho}$ 
do not disturb the KS argument 
\cite{bib:Nagata2}.

The KS theorem states 
one can not construct the realistic functions in the Hilbert space 
formalism of quantum theory in the KS manner.
In more detail, the KS paradox is derived if we assume 
(i) realistic functions exist, 
and 
(ii) the product rule holds.
The product rule is as follows:
when observables $o_1$ and $o_2$ commute, then
\begin{eqnarray}
f(o_1,\omega)f(o_2,\omega)=f(o_1\times o_2,\omega).\label{product}
\end{eqnarray}

Given observables 
$\sigma^a_x$,
$\sigma^a_y$,
$\sigma^a_z$, 
$\sigma^b_x$,
$\sigma^b_y$,
$\sigma^b_z$, 
$\sigma^a_x\sigma^b_x$,  
$\sigma^a_y\sigma^b_y$, 
$\sigma^a_x\sigma^b_y$, 
$\sigma^a_y\sigma^b_x$, and 
$\sigma^a_z\sigma^b_z$, 
we assume that there exist realistic functions
$f(\sigma^a_x, \omega)$,
$f(\sigma^a_y, \omega)$,
$f(\sigma^a_z, \omega)$,
$f(\sigma^b_x, \omega)$,
$f(\sigma^b_y, \omega)$,
$f(\sigma^b_z, \omega)$,
$f(\sigma^a_x\sigma^b_x, \omega)$,
$f(\sigma^a_y\sigma^b_y, \omega)$, 
$f(\sigma^a_x\sigma^b_y, \omega)$,
$f(\sigma^a_y\sigma^b_x, \omega)$, and
$f(\sigma^a_z\sigma^b_z, \omega)$.
Please notice that the possible values of each of realistic functions $f$
are $\pm1$.
In the following, we shall construct the inequality-form-statistical 
KS theorem.
To simplify the discussion,
we assume that (\ref{product}) holds in every hidden variable $\omega$.
(See \cite{bib:Nagata2} for more complicated discussion).
And we shall prove that each of violations of the Ekert inequality and 
of the BBM inequality 
provides explicit sufficient conditions 
for the KS refutation of realism.
Let us consider the three cases.


Case (I):(the case of $\langle\Psi^+|\rho|\Psi^+\rangle>\frac{1}{2}$).
Let $x, y$ be real numbers with
$x, y\in\{-1,+1\}$. Then we have
\begin{eqnarray}
(1+x+y-xy)=\pm 2.\label{CHsimm1}
\end{eqnarray}
The (\ref{CHsimm1}) says
$
U_1(\omega)=1+f(\sigma^a_x\sigma^b_x, \omega)
+f(\sigma^a_y\sigma^b_y, \omega)
-f(\sigma^a_x\sigma^b_x, \omega)
f(\sigma^a_y\sigma^b_y, \omega)\Rightarrow 
U_1(\omega)=\pm2
$
and
\begin{eqnarray}
E_{\rho}(U_1)=\int_{\omega\in\Omega}\mu_{\rho}(d\omega)
U_1(\omega)\leq 2.
\end{eqnarray}
We can factorize two of the terms as
$f(\sigma^a_x\sigma^b_x)=
f(\sigma^a_x)
f(\sigma^b_x)$ and
$f(\sigma^a_y\sigma^b_y)=
f(\sigma^a_y)
f(\sigma^b_y)$.
Further, we have 
$f(\sigma^a_x\sigma^b_y)=
f(\sigma^a_x)
f(\sigma^b_y)$ and
$f(\sigma^a_y\sigma^b_x)=
f(\sigma^a_y)
f(\sigma^b_x)$.
Hence we get 
$f(\sigma^a_x\sigma^b_x)
f(\sigma^a_y\sigma^b_y)=
f(\sigma^a_x\sigma^b_y)
f(\sigma^a_y\sigma^b_x)$
and
\begin{eqnarray}
&&
f(\sigma^a_x\sigma^b_x, \omega)
f(\sigma^a_y\sigma^b_y, \omega)
=
f(\sigma^a_x\sigma^b_y, \omega)
f(\sigma^a_y\sigma^b_x, \omega)\nonumber\\
&&=
f(\sigma^a_z\sigma^b_z, \omega).
\end{eqnarray}
Hence we conclude 
\begin{eqnarray}
&&\int_{\omega\in\Omega}\mu_{\rho}(d\omega)
U_1(\omega)\leq 2\nonumber\\
&&\Leftrightarrow  1+
tr[\rho\sigma^a_x\sigma^b_x]
+tr[\rho\sigma^a_y\sigma^b_y]
-tr[\rho\sigma^a_z\sigma^b_z]\leq 2.\label{KSineq1}
\end{eqnarray}
The violation of the inequality (\ref{KSineq1}) implies 
that one can not introduce the realistic functions satisfying 
the product rule (\ref{product}) in the candidate state $\rho$.
The relation (\ref{KSineq1}) rewritten as
$tr[\rho|\Psi^+\rangle\langle\Psi^+|]\leq \frac{1}{2}$.
Therefore the violation of (\ref{Ekert}), 
because of $\langle\Psi^+|\rho|\Psi^+\rangle>\frac{1}{2}$,
is a sufficient condition for certification of the KS theorem.


Case (II):(the case of $\langle\Psi^-|\rho|\Psi^-\rangle>\frac{1}{2}$).
Note the following relation:
\begin{eqnarray}
(1-x-y-xy)=\pm 2.\label{CHsimm2}
\end{eqnarray}
The (\ref{CHsimm2}) says
$
U_2(\omega)=1-f(\sigma^a_x\sigma^b_x, \omega)
-f(\sigma^a_y\sigma^b_y, \omega)-f(\sigma^a_x\sigma^b_x, \omega)
f(\sigma^a_y\sigma^b_y, \omega)\Rightarrow U_2(\omega)=\pm2
$
and
$E_{\rho}(U_2)\leq 2$.
Similar to the Case (I), we have
$
f(\sigma^a_x\sigma^b_x, \omega)
f(\sigma^a_y\sigma^b_y, \omega)
=
f(\sigma^a_x\sigma^b_y, \omega)
f(\sigma^a_y\sigma^b_x, \omega)
=
f(\sigma^a_z\sigma^b_z, \omega)
$.
Hence we conclude 
\begin{eqnarray}
&&E_{\rho}(U_2)\leq 2\nonumber\\
&&\Leftrightarrow  1-
tr[\rho\sigma^a_x\sigma^b_x]
-tr[\rho\sigma^a_y\sigma^b_y]
-tr[\rho\sigma^a_z\sigma^b_z]\leq 2.\label{KSineq2}
\end{eqnarray}
The violation of the inequality (\ref{KSineq2}) implies 
that one can not introduce the realistic functions satisfying 
the product rule (\ref{product}) in the candidate state $\rho$.
The relation (\ref{KSineq2}) rewritten as
$tr[\rho|\Psi^-\rangle\langle\Psi^-|]\leq \frac{1}{2}$.
Therefore each of violations of (\ref{Ekert}) and of (\ref{BBM}),
because of
$\langle\Psi^-|\rho|\Psi^-\rangle>\frac{1}{2}$,
is a sufficient condition for certification of the KS theorem.


Case (III):(the case of $\langle\Phi^+|\rho|\Phi^+\rangle>\frac{1}{2}$).
Note the following relation:
\begin{eqnarray}
(1+x-y+xy)=\pm 2.\label{CHsimm3}
\end{eqnarray}
The (\ref{CHsimm3}) says
$
U_3(\omega)=1+f(\sigma^a_x\sigma^b_x, \omega)
-f(\sigma^a_y\sigma^b_y, \omega)
+f(\sigma^a_x\sigma^b_x, \omega)
f(\sigma^a_y\sigma^b_y, \omega)\Rightarrow 
U_3(\omega)=\pm2
$
and
$
E_{\rho}(U_3)\leq 2
$.
Similar to the Case (I), we have
$
f(\sigma^a_x\sigma^b_x, \omega)
f(\sigma^a_y\sigma^b_y, \omega)
=
f(\sigma^a_x\sigma^b_y, \omega)
f(\sigma^a_y\sigma^b_x, \omega)
=
f(\sigma^a_z\sigma^b_z, \omega)
$.
Hence we conclude 
\begin{eqnarray}
&&E_{\rho}(U_3)\leq 2\nonumber\\
&&\Leftrightarrow  1+
tr[\rho\sigma^a_x\sigma^b_x]
-tr[\rho\sigma^a_y\sigma^b_y]
+tr[\rho\sigma^a_z\sigma^b_z]\leq 2.\label{KSineq3}
\end{eqnarray}
The violation of the inequality (\ref{KSineq3}) implies 
that one can not introduce the realistic functions 
satisfying the product rule (\ref{product}) in the candidate state $\rho$.
The relation (\ref{KSineq3}) rewritten as
$tr[\rho|\Phi^+\rangle\langle\Phi^+|]\leq \frac{1}{2}$.
Therefore the violation of (\ref{BBM}), 
because of $\langle\Phi^+|\rho|\Phi^+\rangle>\frac{1}{2}$, 
is a sufficient condition for certification of the KS theorem.


Hence, we have, completely, proven that 
each of violations of the Ekert inequality and 
of the BBM inequality 
provides explicit sufficient conditions 
for the KS refutation of realism 
via the existence of the candidate state $\rho$ and 
inequalities (\ref{KSineq1}),
(\ref{KSineq2}), and 
(\ref{KSineq3}).

Assume the inequalities (\ref{KSineq1}) and (\ref{KSineq2}) are valid 
simultaneously, introducing an arbitrary candidate state $\rho$.
Then, we have
$\langle\Psi^+|\rho|\Psi^+\rangle\leq \frac{1}{2}$
and
$\langle\Psi^-|\rho|\Psi^-\rangle\leq\frac{1}{2}$.
Hence the Ekert inequality must be satisfied.
Therefore, the inequalities (\ref{KSineq1}) and (\ref{KSineq2}) give 
a sufficient condition to satisfy the Ekert inequality.
Now, assume the inequalities (\ref{KSineq2}) and (\ref{KSineq3}) are valid
simultaneously.
Then, we have
$\langle\Phi^+|\rho|\Phi^+\rangle\leq \frac{1}{2}$ 
and 
$\langle\Psi^-|\rho|\Psi^-\rangle\leq\frac{1}{2}$.
Hence the BBM inequality must be satisfied.
Therefore, the inequalities (\ref{KSineq2}) and (\ref{KSineq3}) give 
a sufficient condition to satisfy the BBM inequality.
This means that the violation of 
the Ekert and BBM inequalities is impossible 
if the KS type of realism is valid 
in the Hilbert space formalism of quantum theory.
In other words, 
if one certifies the success in each QKD scheme, 
one certifies the KS refutation of 
realism.

Now, assume that 
we abandon the product rule. 
And we introduce locality condition (Bell's locality condition) 
instead of the KS condition (the product rule). 
Bell's locality condition 
is expressed by the fact 
that the realistic functions 
for Alice is independent of 
Bob's ones and vice versa.
When there is it, one can 
construct explicitly a local realistic model for 
given correlation functions \cite{bib:Fine}.
Even in the case, 
the violation of the Ekert and of 
the BBM inequalities is possible as shown below.

Imagine the candidate state $\rho$ is the EPR state 
$|\Psi^-\rangle\langle\Psi^-|$.
Then we have
$tr[\rho\sigma^a_x\sigma^b_x]=-1$,
$tr[\rho\sigma^a_z\sigma^b_z]=-1$,
$tr[\rho\sigma^a_x\sigma^b_z]=0$, and
$tr[\rho\sigma^a_z\sigma^b_x]=0$.
The set of correlation functions is described
with the property that they are reproducible by local realistic theory.
See the following relations 
along with the arguments in \cite{bib:Fine}
\begin{eqnarray}
&&|tr[\rho\sigma^a_x\sigma^b_x]
-tr[\rho\sigma^a_z\sigma^b_z]
+tr[\rho\sigma^a_x\sigma^b_z]
+tr[\rho\sigma^a_z\sigma^b_x]|\leq 2\nonumber\\
&&|tr[\rho\sigma^a_x\sigma^b_x]
+tr[\rho\sigma^a_z\sigma^b_z]
-tr[\rho\sigma^a_x\sigma^b_z]
+tr[\rho\sigma^a_z\sigma^b_x]|\leq 2\nonumber\\
&&|tr[\rho\sigma^a_x\sigma^b_x]
+tr[\rho\sigma^a_z\sigma^b_z]
+tr[\rho\sigma^a_x\sigma^b_z]
-tr[\rho\sigma^a_z\sigma^b_x]|\leq 2\nonumber\\
&&|tr[\rho\sigma^a_x\sigma^b_x]
-tr[\rho\sigma^a_z\sigma^b_z]
-tr[\rho\sigma^a_x\sigma^b_z]
-tr[\rho\sigma^a_z\sigma^b_x]|\leq 2.\nonumber\\
\end{eqnarray}
But, the set of correlation functions produces the violation of
the BBM inequality (\ref{BBM}).

Now, imagine the candidate state $\rho$ is another EPR state 
$|\psi\rangle\langle\psi|$ with 
$|\psi\rangle=\frac{|+^a -^b\rangle+e^{-i(\pi/4)}|-^a +^b\rangle}{\sqrt{2}}$.
Then we have
$tr[\rho a_1 b_1]=1$,
$tr[\rho a_3 b_3]=1$,
$tr[\rho a_1 b_3]=0$, and
$tr[\rho a_3 b_1]=0$.
(Please note
$a_1=\sigma^a_x$, 
$a_3=\sigma^a_y$,
$b_1=\frac{\sigma^b_y+\sigma^b_x}{\sqrt{2}}$, and
$b_3=\frac{\sigma^b_y-\sigma^b_x}{\sqrt{2}}$).
The set of correlation functions is, also, described
with the property that they are reproducible by local realistic theory.
See the following relations,
\begin{eqnarray}
&&|tr[\rho a_1 b_1]
-tr[\rho a_3 b_3]
+tr[\rho a_1 b_3]
+tr[\rho a_3 b_1]|\leq 2\nonumber\\
&&|tr[\rho a_1 b_1]
+tr[\rho a_3 b_3]
-tr[\rho a_1 b_3]
+tr[\rho a_3 b_1]|\leq 2\nonumber\\
&&|tr[\rho a_1 b_1]
+tr[\rho a_3 b_3]
+tr[\rho a_1 b_3]
-tr[\rho a_3 b_1]|\leq 2\nonumber\\
&&|tr[\rho a_1 b_1]
-tr[\rho a_3 b_3]
-tr[\rho a_1 b_3]
-tr[\rho a_3 b_1]|\leq 2.\nonumber\\
\end{eqnarray}
But, the set of correlation functions produces the violation of
the Ekert inequality (\ref{Ekert}).

The explicit difference between the KS condition (the product rule) 
and Bell's locality condition onto the realistic functions $f$ 
in the Hilbert space formalism of quantum theory 
is thus laid bare.

In summary,
we have constructed explicitly 
three inequalities as tests for the KS theorem.
A set of two of those provides a sufficient
condition to satisfy the Ekert inequality.
Another set of two of those provides a sufficient
condition to satisfy the BBM inequality.
Therefore, it turned out that (1) the violation of the Ekert inequality is a 
sufficient condition for certification of the KS theorem, 
and (2) the violation 
of the BBM inequality is, also, 
a sufficient condition for certification of
the KS theorem.
Therefore the success in each QKD protocol 
reveals the nonclassical feature of quantum theory,
in the sense that the KS realism is violated.
This means that quantum observables must be non-commutative
if one wants the success in these EPR-type-QKD protocols 
(Ekert 91 and BBM 92).
Further, it turned out that both (Ekert and BBM) inequalities 
are distillable entanglement witness inequalities.
Thereby we connect the success in each of 
QKD processes into the no-hidden-variables theorem 
and into the confirmation of distillable entanglement.
And we have shown that the violation of 
the Ekert and BBM inequalities in each of QKD protocols 
does not, in general, violate
Bell's local realism.

As the BB84 scheme is essentially equivalent to the BBM scheme, it is very interesting to research the relation between the security of the BB84 
scheme \cite{bib:Shor,bib:Koashi} and 
the KS condition, which  
leads us to stronger refutation of
realism in the formalism of quantum theory
than those revealed by Bell's locality condition.

\acknowledgments

The author would like to 
thank M. Sasaki, M. Takeoka, and M. Koashi for valuable discussions.

\end{document}